\documentclass[aps,nofootinbib,preprint]{revtex4}
\usepackage{graphicx}
\usepackage{amsmath}
\usepackage{amsfonts}
\usepackage{amssymb}
\usepackage{appendix}
\usepackage{mathtools}
\usepackage{comment}
\usepackage{color}
\usepackage{slashed}
\usepackage{subfigure}
\usepackage{setspace}

\begin{document}

\singlespacing

\title{Measuring the Mass of Dark Matter at the LHC}

\author{Andrew C.~Kobach} 
\email[email: ]{akobach@u.northwestern.edu}
\affiliation{Northwestern University, Department of Physics \& Astronomy, 2145 Sheridan Road, Evanston, IL 60208, USA}
\date{\today}

\begin{abstract}
Many methods have been developed for measuring the mass of invisible particles that only use kinematic information available at hadron colliders.  Because a particle is identified by its mass, these methods are critical when distinguishing between dark matter and fake dark matter, where a neutrino or other massless states can mimic a dark-matter signal.  However, the uncertainty associated with measuring the mass of an invisible particle could be so large that it is indistinguishable from a neutrino.  Monte Carlo is used to estimate lower bounds on how heavy an invisible particle must be in order for it to be distinguishable from a massless one at 95\% CL, which we estimate to be $\mathcal{O}$(10 GeV).   This result, to a good approximation, is independent of the way the massive final-state particle is produced.  If there is a light dark-matter particle with mass $\mathcal{O}$(10 GeV), its presence will be difficult to unambiguously identify at the LHC, using kinematic information alone. 
\end{abstract}

\maketitle

\section{Introduction}
\label{intro}

If the experiments at the Large Hadron Collider (LHC) observe sufficient deviation from the standard-model (SM) expectation, then this could be evidence of particles not present in the SM.  These particles' quantum numbers will be intensely investigated, and their masses will be of particular interest.  Because the masses of elementary particles are phenomenological inputs, it is important to develop and utilize methods that can measure the masses of particles produced at a hadron collider.\footnote{In general, the methods used to measure particle masses at hadron and $e^+e^-$ colliders would not be identical, because the initial-state energy of event is unknown at a hadron collider}
 
There are two strategies for measuring masses at a hadron collider.  The first method relies on measuring decay rates, lifetimes, etc., whose values, in general, depend on the masses of the particles in the event, e.g., measuring the mass of the muon by measuring its lifetime.  However, this method is uncommon, in general, because it requires information about matrix elements.  The second method relies on directly measuring the 4-vectors of the particles in an event.  This is a preferred means to measure masses, because it is independent of a matrix element.  For example, one can measure the mass of the $Z$ boson using dilepton events, where the 3-vectors of the leptons are measured by the detector, and since the leptons are approximately massless, their 4-vectors are inferred.  However, in other kinds of events at hadron colliders, not all 4-vectors components can be directly measured or inferred, and measuring masses can be nontrivial. 

To avoid relying on detector-related variables, e.g.,~charged tracks, depositions in the electromagnetic or hadronic calorimeters, etc., this analysis is framed in terms of the known and unknown components of the 4-momenta in the event.  The scenario of interest is how to measure the mass of a collider-stable particle without information regarding its energy.  For example, consider a collection of events at the LHC where dark matter is produced in the final state from the decay of a single parent particle with unknown mass.  Only a subset of the dark matter's 3-momentum can be measured, and its energy is, in general, unknown.  Measuring the mass of dark matter produced at a hadron collider proves to be a unique challenge, and many methods have been developed, which, in principle, can do so, e.g., as those described in Refs.~\cite{Cho:2007qv, Gripaios:2007is, Barr:2007hy, Ross:2007rm, Nojiri:2007pq, Nojiri:2008hy, Tovey:2008ui, Cheng:2008mg, Barr:2008ba, Kersting:2008qn, Cho:2008tj,  Burns:2008va, Barr:2008hv, Kersting:2009ne, Burns:2009zi, Han:2009ss, Webber:2009vm, Matchev:2009fh, Cho:2009wh, Kim:2009si, Matchev:2009ad, Konar:2009wn, Autermann:2009js, Konar:2009qr, Cohen:2010wv, Kang:2010hb, Cheng:2011ya, Cho:2012er, Han:2012nm, Han:2012nr}.  These methods assume a particular event topology in which dark matter is produced and demonstrate that its mass (and the mass of its parent) can be experimentally measured using only kinematic information available in the event. 

There is considerably less work in the present literature concerning {\it how well} the mass of dark matter can be measured.  When measuring the mass of an invisible particle at the LHC, could the error bar be so big that the measurement is not meaningful?  There are models of fake dark matter, where a missing-energy signal is due instead to the anomalous production of neutrinos~\cite{Chang:2009dh}.  If dark matter is light, say 1--10 GeV, then it may become difficult to distinguish between these scenarios without relying on information about the form of the matrix element.  

We attempt to estimate a lower bound on how heavy dark matter must be in order for it to be distinguishable at 95\% CL from a massless state, which is a value we call $m_\chi^\text{min}$.  Directly measuring the mass of dark matter involves the convolution of two independent challenges.  The first is measuring the mass of a collider-stable particle when only its 3-momentum, and not its energy, is measured and the mass of its parent particle is unknown.  The second challenge is that only a subset of its 3-momentum is reconstructed.  In order to estimate a lower bound on the value of $m_\chi^\text{min}$, we ignore the latter challenge, because it presupposes the former.  By doing so, we permit ourselves to have information regarding all the 3-momentum components, the value of $m_\chi^\text{min}$ must be equal to or less than its value if only a subset were reconstructed.  This will allow the value of $m_\chi^\text{min}$ to be roughly independent of the way the dark matter particle was produced, which we will explore later. 

We describe a method in Section~\ref{section2} to measure the mass of a final-state particle, $\chi$, produced from a parent particle with unknown mass, $A$, and a massless sibling, $B$, i.e., $A\rightarrow B\chi$.  In general, $A$ could be an intermediate particle, a part of a larger decay topology.  To further underestimate the value $m_\chi^\text{min}$, we assume there is no background contamination, that all 3-momenta can be reconstructed with a very optimistic resolution, there is no combinatorial ambiguity associated with the event, and all particles are on-shell.  We simulate the production of $A$, and its subsequent decay, for event topologies that resemble $t\bar{t}$ production and $WW$ production, as described in Section~\ref{section3}.  The method in Section~\ref{section2} is employed to simultaneously measure the value of $m_\chi$ and $m_A$ and estimate the $m_\chi^\text{min}$ as a function of $m_A$.   We find that $m_\chi^\text{min}$ has a value for $\mathcal{O}$(10 GeV), and this does not have strong dependence on the event topology.   

Some may find it compelling that the results from the CoGeNT~\cite{Aalseth:2010vx}, DAMA/LIBRA~\cite{Bernabei:2008yi,Bernabei:2010mq}, CDMS~\cite{Agnese:2013rvf}, and CRESST-II~\cite{Angloher:2011uu} experiments are consistent with a light dark matter particle of mass $\approx$ 10 GeV.  If such a light dark-matter candidate were produced at a hadron collider, then our results suggest, independent of the event topology, that information other than the dark matter's mass would be required to identify it at the LHC.

\section{Measuring the Mass of Final-State Particles }
\label{section2}

Consider the kinematics of the two-body decay, $A\rightarrow B\chi$.  The 3-momenta of the daughter particles, $\vec{\bf p}_B$ and $\vec{\bf p}_\chi$, are fully reconstructable, and the values of $m_\chi$ and $m_A$ are in general unknown.  One can define an ansatz for the value of $m_\chi$, called $\tilde{m}_\chi$.  The squared invariant mass of this system, $M^2$, can be written as a function of $\vec{\bf p}_B$, $\vec{\bf p}_\chi$, and $\tilde{m}_\chi$ as
\begin{equation}
\label{invmass}
M^2(\tilde{m}_\chi) \equiv \tilde{m}_\chi^2 + m_B^2 + 2\left( |\vec{\bf p}_B| \sqrt{\tilde{m}_\chi^2 + \vec{\bf p}_\chi} - \vec{\bf p}_B \cdot \vec{\bf p}_\chi \right).
\end{equation}
From here, we assume $A$ is on-shell, and for simplicity, $m_B = 0$.   Given a collection of these events in the center-of-mass (CM) frame of $A$, a histogram of $M$, for any value of $\tilde{m}_\chi$, will resemble a delta function, and, in particular, $M(\tilde{m}_\chi = m_\chi) = m_A$.  On the other hand, if $A$ is boosted in a different direction relative to the CM frame for each event, and assuming perfect resolution of $\vec{\bf p}_B$ and $\vec{\bf p}_\chi$, then a histogram of $M$ will resemble a delta function only if $\tilde{m}_\chi = m_\chi$.  Otherwise, the distribution of $M$ will have some spread.  

\begin{figure}[htbp]
\begin{center}
\includegraphics[width=0.55\textwidth]{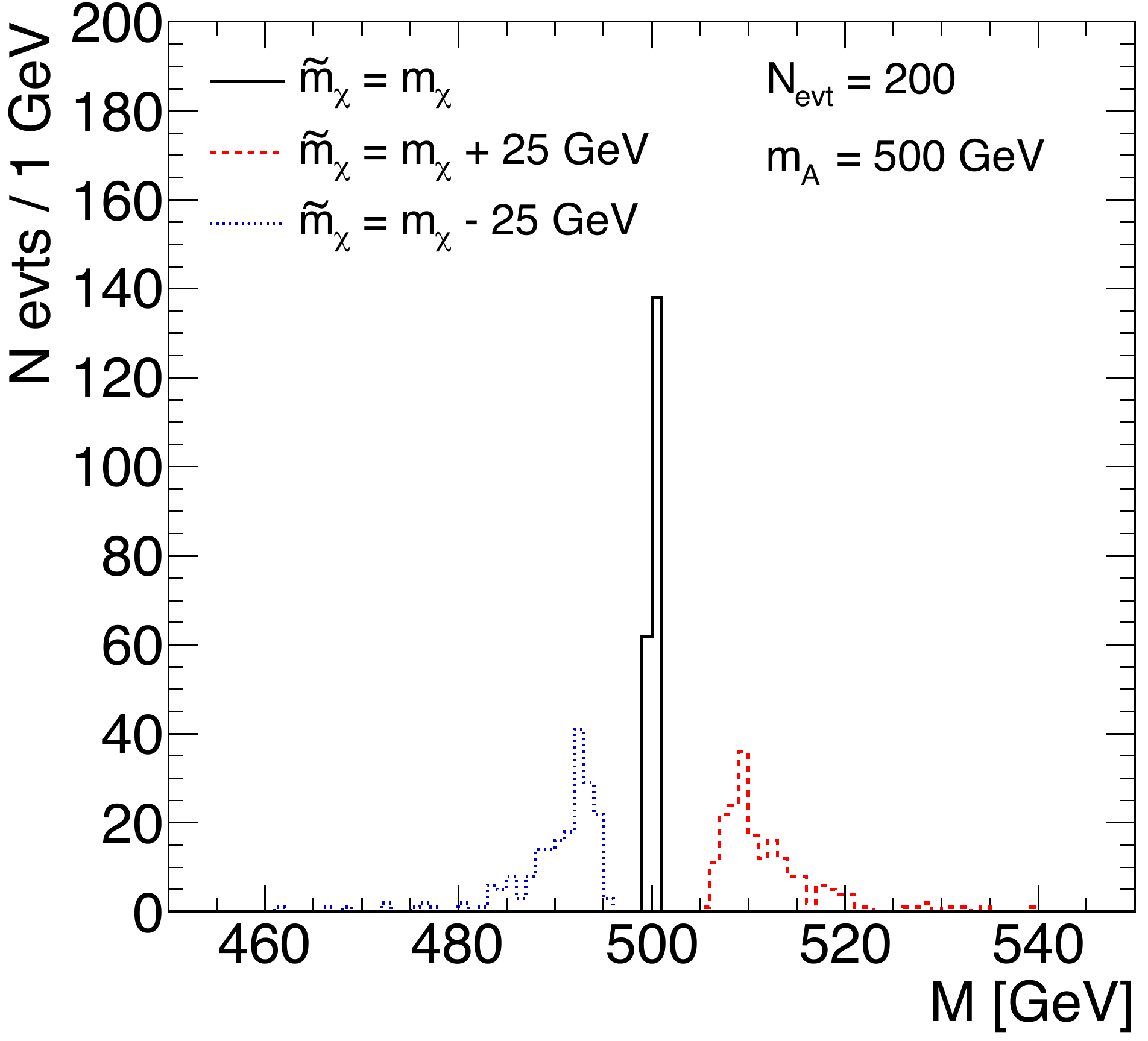}
\caption{Histograms of $M$, as defined in Eq.~(\ref{invmass}), for the decay $A\rightarrow B\chi$, where $m_B=0$, $m_A$ = 500 GeV, and the 3-momentum of the final-state particles are fully reconstructable.  The value of $\tilde{m}_\chi$ is varied to demonstrate that the distribution of $M$ is the narrowest when $\tilde{m}_\chi$ is equal to the physical mass, $m_\chi$.}
\label{diffansatz}
\end{center}
\end{figure}

To demonstrate this effect, we simulate 200 MSSM disquark events (the squark decays to a quark and the LSP) simulated with {\sc madgraph5}~\cite{Alwall:2011uj} for LHC events at $\sqrt{s} = 14$ TeV, where the squark and LSP masses are 500 GeV and 100 GeV, respectively.  The value of $M$ is calculated using Eq.~(\ref{invmass}) for each event, given value of $\tilde{m}_\chi$.  Fig.~\ref{diffansatz} shows the histograms when $\tilde{m}_\chi$ are 75 GeV, 100 GeV, and 125 GeV, and the shape of $M$ is the narrowest when $\tilde{m}_\chi = m_\chi$.  
When boosting from the CM frame to the lab frame, the components of $\vec{\bf p}_\chi$ that are parallel to the boost direction will mix with the energy of the $\chi$ particle, which contains information about $m_\chi$.  If $\tilde{m}_\chi \neq m_\chi$, the value of $M$ in the lab frame will depend on how $A$ was boosted, which implies it is no longer a Lorentz invariant.  Given only the measurements of $\vec{\bf p}_B$ and $\vec{\bf p}_\chi$, one can, in principle, simultaneously measure $m_\chi$ and $m_A$ by finding the value of $\tilde{m}_\chi$ for which the distribution of $M$ is the most narrow.  This will remain true when $\vec{\bf p}_B$ and $\vec{\bf p}_\chi$ are subject to finite experimental resolution.   Measuring $m_A$ and $m_\chi$ depends mostly on whether or not $A$ is boosted randomly among a collection of events and less on its precise momentum distribution.  For this reason, we can naively expect that this method will not be strongly sensitive to the way $A$ was produced.

\section{Analysis}
\label{section3}
\begin{figure}[htbp]
\begin{center}
\subfigure[]{\label{type1}\includegraphics[width=0.4\textwidth]{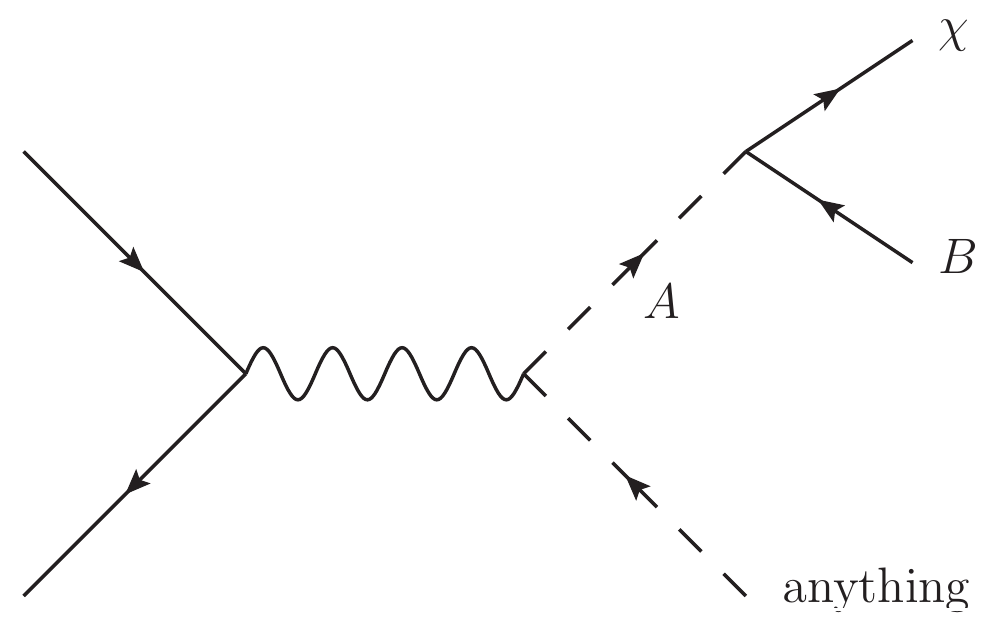}}\\
\subfigure[]{\label{type2}\includegraphics[width=0.5\textwidth]{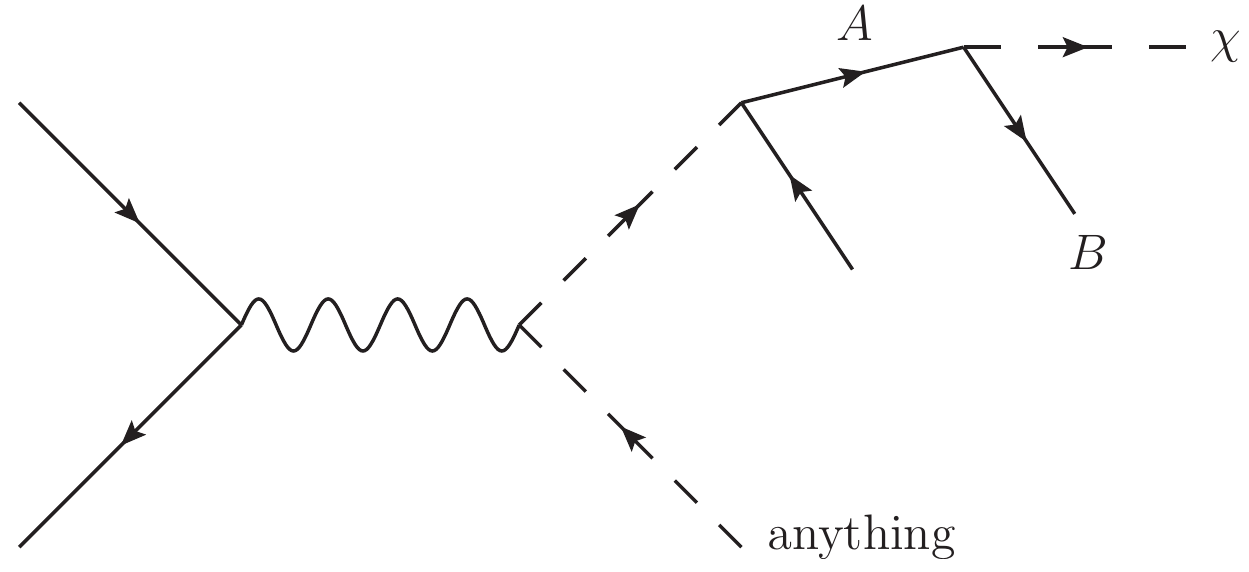}}
\caption{(a) Type-I and (b) Type-II decay topologies.  The analysis is insensitive to whether the particles $A$, $B$, or $\chi$ are bosons or fermions.}
\label{decaytypes}
\end{center}
\end{figure}

We study two types of event topologies in which dark matter could be produced, called Type-I and Type-II, which can be found in Figs.~\ref{type1} and \ref{type2}, respectively.  The method described in Section~\ref{section2} is used to measure $m_\chi$, which is treated as a visible final-state particle.  The magnitude of the measurement's uncertainty will depend on the number of signal events, the value of $m_A$, and how $A$ is produced in a larger decay topology.  In particular, we choose $N=$ 200, 500, and 1000 signal events and values of $m_A$ between 100 GeV and 1 TeV.  These values of $N$ were chosen because they would lead to a clear experimental signal.  Varying $N$ by a factor of five allows us to see how the results change with large changes in statistics.

para

Some types of decay chains in the MSSM are well suited for this analysis.  To simulate Type-I decays,  pseudo-data events of $pp\rightarrow \tilde{q} \bar{\tilde{q}}$ are simulated, with the MSSM {\sc madgraph5} package at $\sqrt{s} =$ 14 TeV, where one of the squarks decays to a quark and the LSP.  Here, the squark can be thought of as $A$, the final-state quark as $B$, and the LSP as $\chi$.  To simulate Type-II decays topologies, pseudo-data events of $pp\rightarrow \tilde{g} \tilde{g}$ are generated, where at least one of the gluinos decays as $\tilde{g}\rightarrow \tilde{q} \bar{q}$, and the squark subsequently decays to a quark and the LSP.  For these events, the mass of the gluino is chosen to be 2 TeV.  Note that our method for measuring the mass of the invisible particle is, in principle, insensitive to whether $A$, $B$, or $\chi$ are  fermions or bosons.

We minimize, as much as possible, the magnitude of the uncertainty associated with measuring $m_\chi$.  In particular, there is no background contamination in the signal sample, $A$ is on-shell, and there is no combinatorial ambiguity associated with identifying $B$ and $\chi$.  We assume, very optimistically, that the magnitude of the 3-momentum of $B$ and $\chi$, $|\vec{\bf p}_B|$ and $|\vec{\bf p}_\chi|$, respectively, smear similarly to electrons at the CMS experiment, according to the parametrization found in Ref.~\cite{Chatrchyan:2009qm},\footnote{This smearing is based off of calorimeter performance.  While in traditional parlance the energy of the electron is smeared, we must generalize this to mean the magnitude of the momentum, because now the final-state particle is massive.} 
\begin{equation}
\label{res}
\frac{\sigma^{e^\pm}_{|\vec{\bf p}|}}{|\vec{\bf p}|} = \frac{0.028}{\sqrt{|\vec{\bf p}|}} \oplus \frac{0.0415}{|\vec{\bf p}|} \oplus 0.003.
\end{equation}
The magnitude of this smearing induces about a 2 GeV Gaussian width of $M$ when $\tilde{m}_\chi = m_\chi$. 

For $N$ pseudo-data events of Type-I or Type-II decays, the invariant mass, $M$, as defined in Eq.~(\ref{invmass}), is reconstructed for a given value of $\tilde{m}_\chi$, $m_A$, and $m_\chi$.  This distribution is then fitted with a Gaussian distribution, the width of the fit is recorded, and the procedure is performed again for a different value of $\tilde{m}_\chi$.\footnote{Distributions other than Gaussians were used to fit the histogram of $M$, and the results did not significantly change. When $\tilde{m}_\chi$ is close to $m_\chi$, the convolution of the Gaussian smearing of the momentum and the widening effects shown in Fig.~\ref{diffansatz} is still approximately Gaussian.  Additionally, the variance of the distribution can be used, which yields almost identical results.}   The value of $\tilde{m}_\chi$ for which the Gaussian width of $M$ has the smallest value is the best estimate for the value of $m_\chi$, called $\tilde{m}_\chi^*$.  This procedure is performed for 2,000 pseudo-experiments, each with $N$ pseudo-data events, resulting in a distribution of 2,000 values of $\tilde{m}_\chi^*$, centered around the physical value of $m_\chi$.  This distribution is integrated and the smallest value of $m_\chi$ for which the distribution of $\tilde{m}_\chi^*$ is not excluded from zero at 95\% CL is the value of $m_\chi^\text{min}$.   

\begin{figure}[htbp]
\begin{center}
\subfigure[]{\label{}\includegraphics[width=0.55\textwidth]{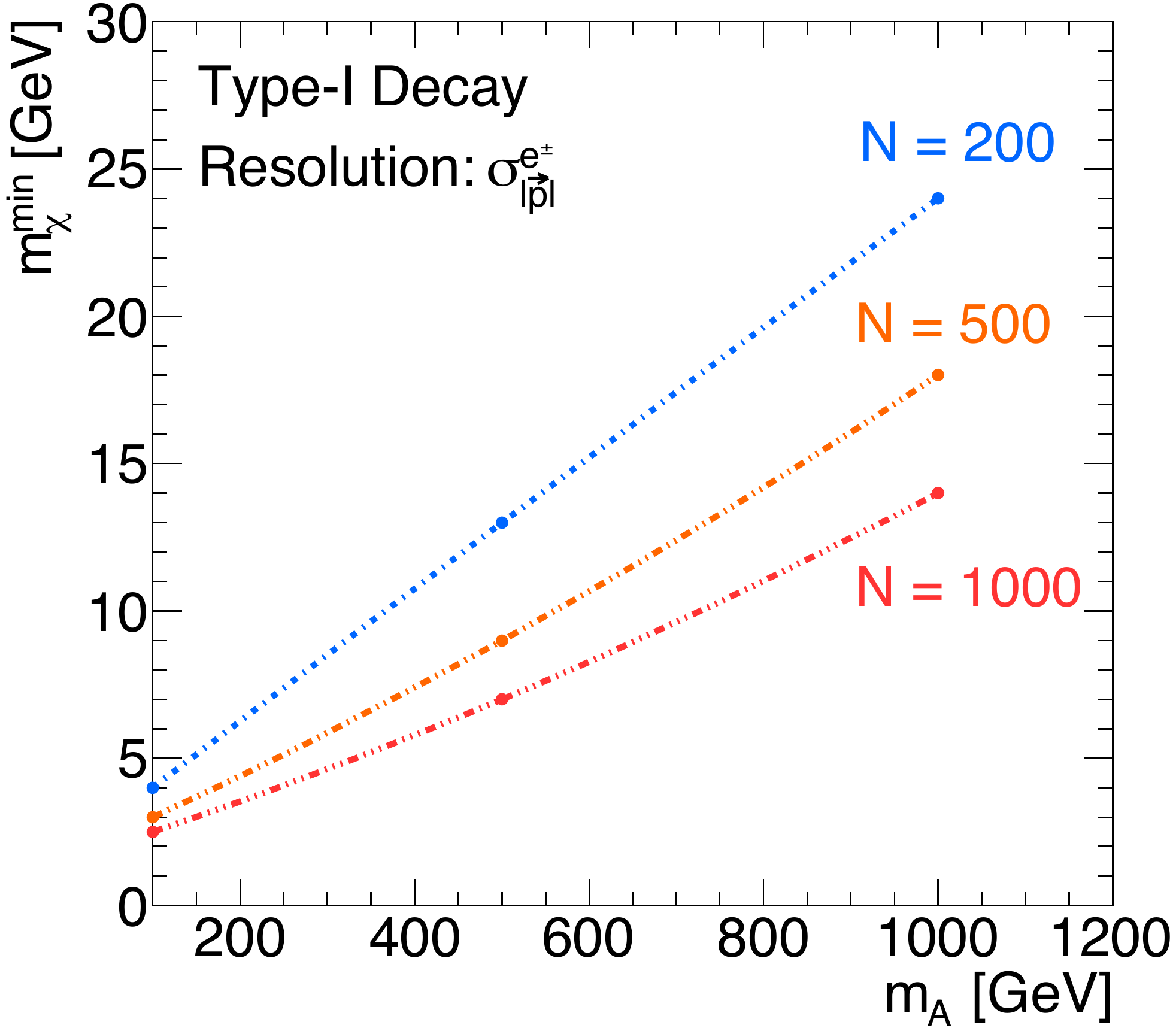}}
\subfigure[]{\label{}\includegraphics[width=0.55\textwidth]{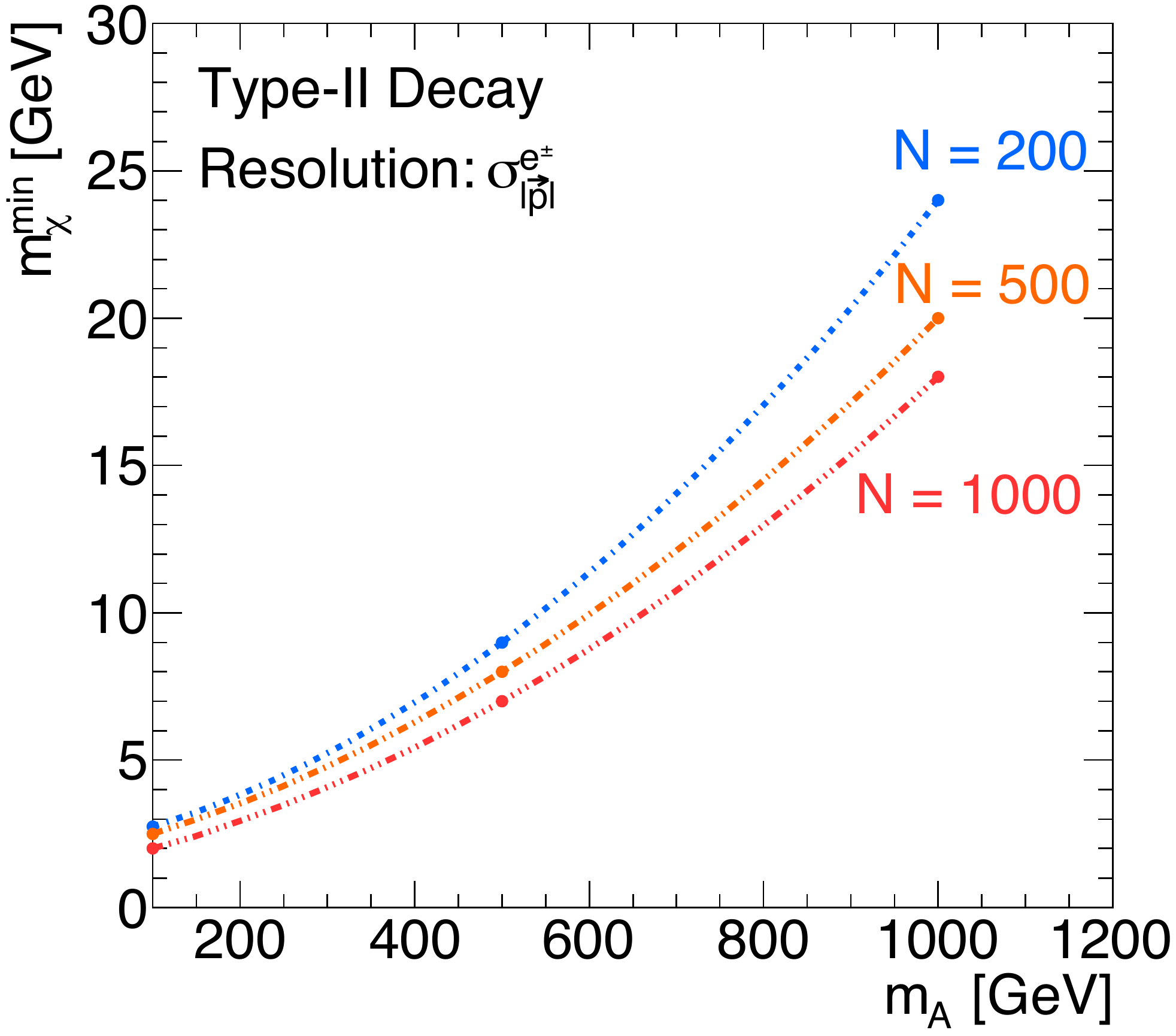}}
\caption{Results for $m_\chi^\text{min}$, as a function of $m_A$, with the momentum resolution as found in Eq.~(\ref{res}), for (a) Type-I and (b) Type-II decay topologies.  The blue dash-dot, the orange dash-dot-dot, and red dash-dot-dot-dot lines correspond to $N$ = 200, 500, and 1000 signal events, respectively. }
\label{mchimin_sigma}
\end{center}
\end{figure}

\begin{figure}[htbp]
\begin{center}
\subfigure[]{\label{}\includegraphics[width=0.55\textwidth]{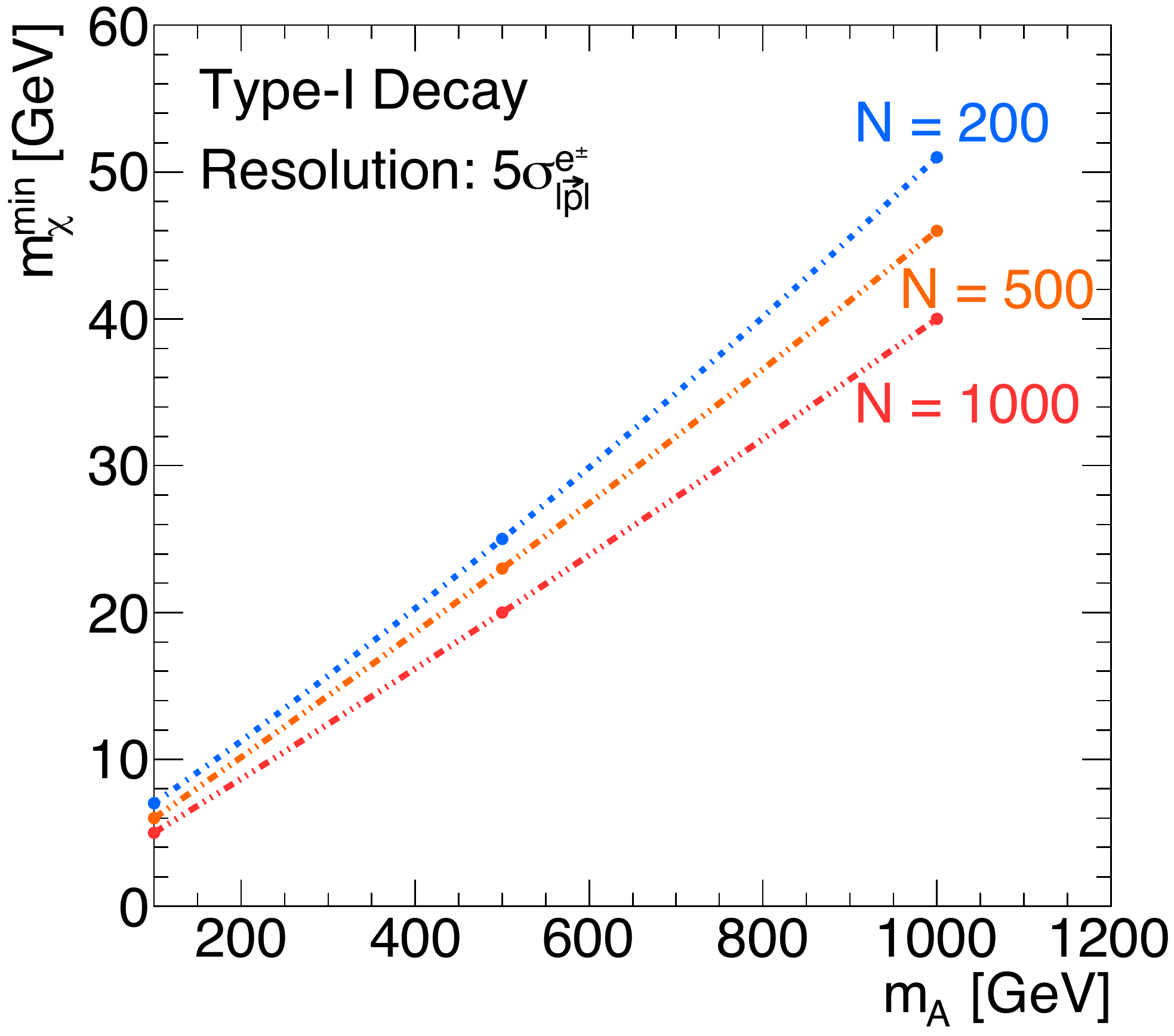}}
\subfigure[]{\label{}\includegraphics[width=0.55\textwidth]{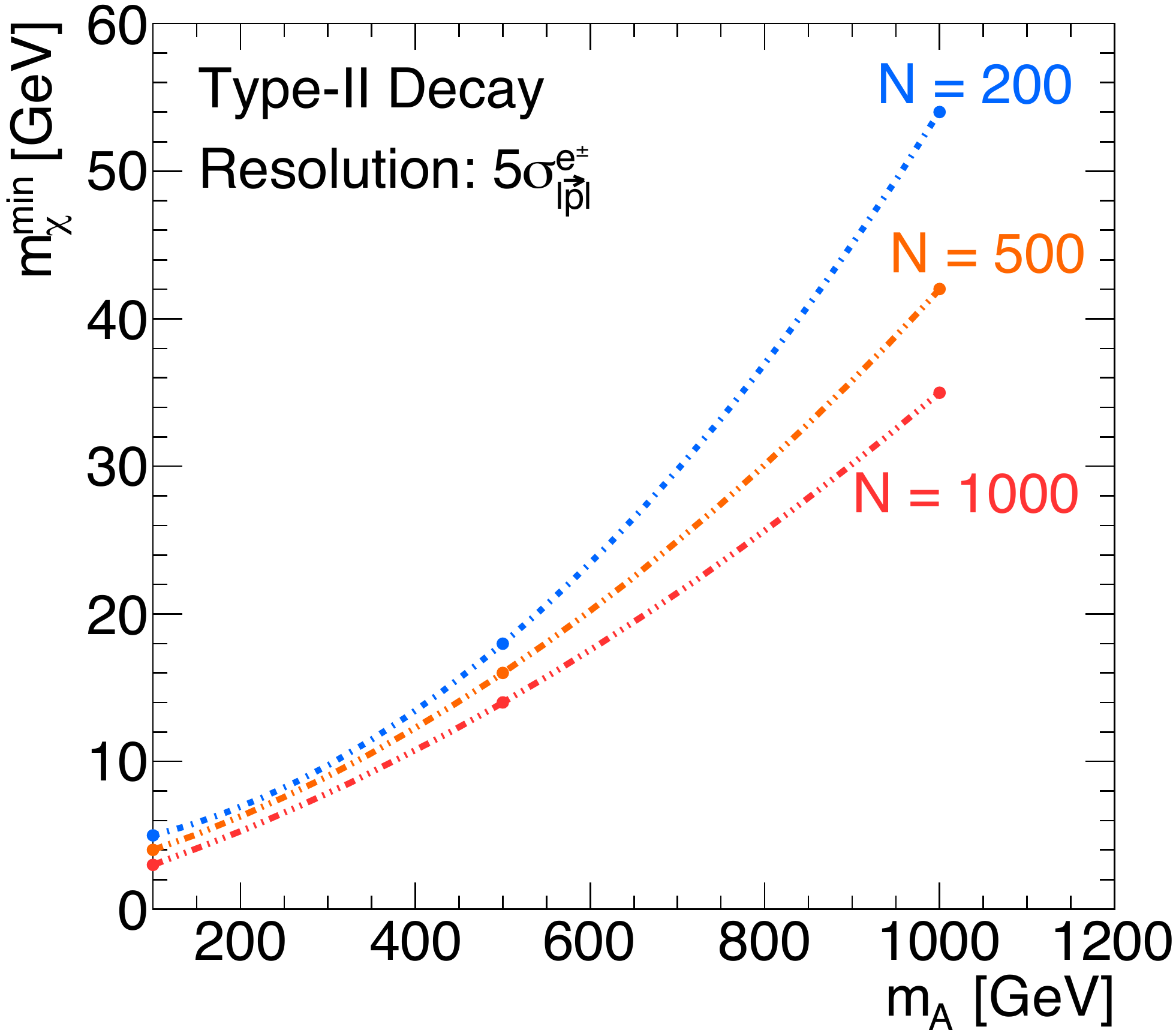}}
\caption{Results for $m_\chi^\text{min}$, as a function of $m_A$, with five times worse momentum resolution as found in Eq.~(\ref{res}), for (a) Type-I and (b) Type-II decay topologies.  The blue dash-dot, the orange dash-dot-dot, and red dash-dot-dot-dot lines correspond to $N$ = 200, 500, and 1000 signal events, respectively. }
\label{mchimin_5sigma}
\end{center}
\end{figure}

Results for the value of $m_\chi^\text{min}$ can be found in Fig.~\ref{mchimin_sigma}, for Type-I and Type-II decays and $N$ = 200, 500, and 1000, with values of $m_A$ between 100 GeV and 1 TeV.   To demonstrate how the values of $m_\chi^\text{min}$ change when the resolution is made less optimistic, the analysis is repeated where $\vec{\bf p}_\chi$ and $\vec{\bf p}_B$ have five times worse resolution:~$5\times\sigma^{e^\pm}_{|\vec{\bf p}|}$.   These results can be found in Fig.~\ref{mchimin_5sigma}.  In general, the value of $m_\chi^\text{min}$ increases as the value of $m_A$ increases, since the momentum of $A$ becomes smaller in magnitude as its mass increases.  The value of $m_\chi^\text{min}$ scales linearly with $m_A$ for Type-I topologies and scales nonlinearly in Type-II topologies.  While the shapes qualitatively differ, the results for $m_\chi^\text{min}$ are quite similar in magnitude for both Type-I and Type-II event topologies.  This was expected, since both $\vec{\bf p}_B$ and $\vec{\bf p}_\chi$ are fully reconstructed, and the ability for one to simultaneously measure $m_A$ and $m_\chi$ depend on $A$ being boosted differently in each event, relative to its CM frame, which the Type-I and Type-II topologies share.  
For both resolutions and decay topologies, as the value of $N$ is increased, for a fixed value of $m_A$, the value of $m_\chi^\text{min}$ scales roughly as the inverse cube-root of the increase of statistics. 
The values for $m_\chi^\text{min}$ are interpreted as lower bounds on how heavy the visible final-state particle must be in order for it to be distinguished from an effectively massless particle.  Consequently, they also serve as lower bounds on how heavy dark matter must be in order to determine it has a nonzero mass at 95\% CL, using only kinematic information.

\section{Conclusion}
\label{section4}
If one accepts that the signals from the CoGeNT~\cite{Aalseth:2010vx}, DAMA/LIBRA~\cite{Bernabei:2008yi,Bernabei:2010mq}, CDMS~\cite{Agnese:2013rvf}, and CRESST-II~\cite{Angloher:2011uu} experiments are due to a dark matter species with mass $\approx$ 10 GeV, then it is possible that the particle could manifest itself at the LHC in events with an excess of missing energy.   However, many new physics scenarios can also give rise to events at the LHC with missing-energy signals, in particular, anomalous neutrino production~\cite{Chang:2009dh}.  A model-independent method to distinguish dark matter production from anomalous neutrino production is to directly measure the mass of the particle associated with the missing energy.   However, due to experimental limitations, it would be a challenge to distinguish light dark matter from other invisible particles with effectively zero mass.  

By estimating the uncertainty associated with measuring the mass of a final-state particle, we estimated lower limits of how heavy the dark matter must be in order for it to be distinguishable from a massless particle at 95\% CL, which we call $m_\chi^\text{min}$.  We assume that dark matter, $\chi$, has a single sibling, $B$, both of which are produced from a single parent particle of unknown mass, i.e., $A\rightarrow B\chi$.  In general, $A$ can be a part of a larger even topology, as shown in Fig.~\ref{decaytypes}.  Many assumptions are made that lead to underestimating the value of $m_\chi^\text{min}$.  We assume these events have no background contamination, the parent particle is on-shell, and there is no combinatorial ambiguity associated with the identification of $B$ and $\chi$.  To further underestimate the value of $m_\chi^\text{min}$, we allow $\chi$ to be visible, i.e., its 3-momentum is completely reconstructible.  By doing this, we permit ourselves to have access to information that is not available when actual dark matter is produced at a hadron collider.  This allows our results to be roughly independent of the larger event topology. 

The method to measure $m_\chi$, as described in Section~\ref{section2}, relies on the parent particle, $A$, being boosted differently for every event between its CM frame and the lab frame.  We investigate different topologies for the production of the parent particle, e.g., Type-I and Type-II decay topologies, as shown in Fig.~\ref{decaytypes}.  While $m_\chi^\text{min}$ does depend on how $A$ is produced, the magnitude for $m_\chi^\text{min}$ is similar for Type-I and Type-II decay topologies, as seen in Figs.~\ref{mchimin_sigma} and \ref{mchimin_5sigma}.  The value for $m_\chi^\text{min}$ increases as $m_A$ increases, because heavier particles tend to have less momentum, which decreases the sensitivity to the mass of the final-state particles, as described in Section~\ref{section2}.  

Upon first glance, the method described in Section~\ref{section2} to measure the mass of a final-state particle seems sufficiently different from $M_{T2}$-based methods.  A $M_{T2}$-based method can be used for a $t\bar{t}$-like decay topology, where two identical decay chains produce a pair of dark-matter particles~\cite{Burns:2008va}.  In this scenario, the mass of the invisible particles is determined by tracking how the $M_{T2}$ endpoint changes as a function of the input ansatz mass.  The method used in this analysis, on the other hand, measures the mass of a final-state particle by measuring the value of the input ansatz mass for which an invariant mass distribution is the narrowest, not how the invariant mass changes with a function of the ansatz.  It is reasonable to suspect that this method and $M_{T2}$-based methods might give different results for $m_\chi^\text{min}$.  However, as shown in Ref.~\cite{deGouvea:2012ez}, for dark matter specifically produced in a $t\bar{t}$-like topology, it is difficult to kinematically distinguish dark matter and neutrinos if the dark matter has a mass below $\mathcal{O}(\text{10 GeV})$, even if the $M_{T2}$ endpoints can be measured with an optimistic precision of 1 GeV, which agrees with experimental results~\cite{Chatrchyan:2013boa}.  As expected, the estimates for the lower bound of measurable dark-matter mass in this analysis are indeed lower than those found in Ref.~\cite{deGouvea:2012ez}. 

Since the values of $m_\chi^\text{min}$ in this analysis are underestimated by assuming the 3-momentum of $\chi$ can be fully reconstructed,  the values of $m_\chi^\text{min}$ would increase if only a subset of the momentum is known, as is the case with invisible particles.  The values of $m_\chi^\text{min}$ in this analysis can be considered as strict lower bounds on how heavy dark matter must be in order to distinguish it from a massless state.  Because particles are identified by their masses, we can expect that if dark matter is light, i.e., $\mathcal{O}(\text{10 GeV})$, as hinted by some direct-detection experiments, it will be  difficult to unambiguously identify its presence at the LHC, using only kinematic information.

\begin{acknowledgments}
The author is grateful to Andr\'e de Gouv\^ea, Jennifer Kile, KC Kong, and Andy Kubik  for useful conversations and feedback.  ACK is supported in part by the Department of Energy Office of Science Graduate Fellowship Program (DOE SCGF), made possible in part by  the American Recovery and Reinvestment Act of 2009, administered by ORISE-ORAU under contract no.~DE-AC05-06OR23100.
\end{acknowledgments}

\bibliography{bib}{}

\end{document}